\documentstyle[aps,epsf]{revtex}

\tightenlines
\newcommand{\infig}[2]{\begin{center}\mbox{\epsfxsize #2
\epsfbox{#1}}\end{center}}

\begin{document}
\title{Long-distance Bell-type tests using energy-time entangled photons}
\author{W.Tittel, J.Brendel, N.Gisin, and H.Zbinden}
\address{University of Geneva, 
         Group of Applied Physics, 
         20,Rue de l'Ecole de Med\'ecine, 
         CH-1211 Geneva 4, Switzerland\\
         email: wolfgang.tittel@physics.unige.ch}

\date{\today}

\maketitle

\begin{abstract}
Long-distance Bell-type experiments are presented.
The different experimental challenges and their solutions in order to 
maintain the strong quantum 
correlations between energy-time entangled photons over more than 10 
km are reported and the
results analyzed from the point of view of tests of fundamental physics 
as well as from the more 
applied side of quantum communication, specially quantum key 
distribution. Tests
using more than one analyzer on each side are also presented.

\end{abstract}

PACS Nos. 3.65BZ, 3.67.Dd 42.81.-i


\section{Introduction}
     
Entanglement, the possibility for a quantum system composed of 
several particles to be in a 
definite state while each single particle is in a mixed state, 
is one of the most interesting and puzzling predictions of quantum 
mechanics. The history goes back 
to 1935. Starting from the perfect 2-particle correlations predicted 
for entangled states, 
Einstein, Podolsky and Rosen argued that quantum theory is not 
complete \cite{EPR}. 
In 1964, Bell demonstrated that the attempt to complete the theory with 
so called hidden variables 
and maintaining the locality condition leads to statistical predictions 
for measurements along 
nonorthogonal bases 
which differ from those given by standard quantum theory \cite {bell64}.
Tests of the so called Bell inequalities have been made again and again 
\cite{belltests,aspect82,aspectswitch,Tapster94,longdistbell} in 
order to show more and more clearly that the quantum-correlations can not 
be explained by local hidden 
variables theories (LHVT). 
Today, most physicists are convinced that a future loophole-free test 
\cite {loopholes}
will definitely demonstrate that nature is indeed nonlocal. 

However, there is still interest in new experiments, the motivation is 
threefold. A first aim in 
future tests is to close the remaining loopholes. For instance, the fact 
that the detected pairs 
of particles form only a small
and possibly biased subensemble of the created pairs, the so-called 
detection loophole \cite{detectionloophole},
is under 
investigation at groups in Los Alamos \cite{LA}, Texas \cite{Texas} and 
Paris \cite{Paris}. 
The locality loophole, 
based on the assumption that the properties of the correlated particles 
might be predetermined
by the settings of the analyzers \cite{localityloophole}, 
has been examined by Aspect et al in 1982 \cite{aspectswitch}. 
Yet, the fast changing of the settings in order to prevent 
any influence on the photon pairs within the frame of special relativity 
has been criticized as not being
really random \cite{zeilingercritic}. 
Recently an experiment has been realized in Innsbruck in order to close 
remaining doubts on the locality 
loophole \cite{innsbruck98}. Our results with more than one analyzer on 
each side can also be
considered in this context (see section V.A).

A second motivation for Bell-type tests is evolving from a recent proposal 
to use entangled particles 
for a test of relativistic 
nonlocality (or multisimultaneity), 
an alternative quantum theoretical description of nature which unifies 
nonlocality and 
relativity of simultaneity \cite {RNL}. 
Such a test requires a large spatial separation between the different 
parts of the experiment. Hence it requires that the quantum correlations 
are maintained
even when separating the particle over scales larger than the usual 
laboratory ones.

Third, besides these roles as candidates for tests of fundamental physics, 
entangled particles lie at the 
heart of the new field of quantum information processing (or quantum 
communication) 
which has evolved in rather big steps during the last years 
\cite {physworld}. Its characteristics 
is to turn quantum conundrums into potentially useful processes which can 
not be achieved using classical 
physics. 
One of the most promising results of this new field is quantum key 
distribution \cite {physworld,ScieAmerican}, often also referred to as 
quantum cryptography, a way to 
establish a secret key between two parties which can be used afterwards 
to code and decode a message.
The quantum mechanical law that a measurement of an unknown system will 
in most of the cases disturb
the system is exploited here to reveal an 
eavesdropper: if none of the transmitted bits, coded in nonorthogonal 
states, have been disturbed, 
no unlegitimate third person has tried to listen 
in. Other examples 
for quantum information processing are dense coding \cite {densecoding} 
(the possibility to send more than 
one bit of classical 
information encoded in a single quantum bit) and teleportation 
\cite {teleportation}
(transmission of an arbitrary quantum state from one particle to 
another one). Algorithms to factorize large 
numbers with a quantum computer are known and are much faster than 
those, known for classical computers 
\cite {qcomputer}. The key word of the whole field of quantum information 
processing is entanglement. 
Two particle entanglement is required for dense-coding and teleportation 
and for some schemes 
of cryptography \cite{ekert91}, entanglement of thousands of particles is 
needed for quantum computers. Hence 
the whole field relies on the existence of quantum nonlocality and on the 
fact that environment-induced 
decoherence \cite{zurek91} (and spontaneous collapses 
\cite{furry36,GRW86}, if it exists) 
can be kept small or can be prevented
for a sufficiently long time or distance.

In 1997 and 1998, we performed two series of experiments in order to 
examine whether energy-time entanglement is robust enough to be really 
exploited for the motivations mentioned in the last two paragraphs 
\cite{notthesame}.
The aim of this article is to draw a concluding line under these 
Bell-type tests of quantum correlations over more than 10 kilometers, 
and to 
provide the reader with more information about experimental requirements 
than has been published in short letters \cite{longdistcorrel,longdistbell}.
As the results from the 1997 experiment \cite{longdistcorrel} are 
confirmed in the 
1998 one \cite{longdistbell}, we will mostly focus on the latter 
experiment as it is 
altogether more complex and uses more
advanced technology. However, interesting experimental solutions, chosen 
in the 1997 experiment 
will be mentioned as well.

The outline of this paper is the following: 
After this introduction we will briefly present the 
theoretical background, for tests of Bell inequalities as well as for 
quantum cryptography (section II). 
Then, we will describe the experimental setup in section III. 
This part is divided into several sub-sections, each one focussing 
on a special detail: two-photon source, interferometers, dispersion in 
optical fibers, photon detectors, 
transmission of results of measurements, and measurement of correlations.
Next, section IV, reports on the results. 
In addition to Bell tests implying one analyzer for each of the correlated 
particles, 
results for an experiment with three analyzers, two on one end and the 
third at the other end 
(10 km away) will be presented. Finally we will report on new data 
obtained in a laboratory 
(short distance) experiment 
using four analyzers, two on each side of the source. In the latter 
experiments, the two nearby 
devices analyse the incoming
photons randomly, the choice being made by a passive beam splitter. 
These setups enable to test directly 
the CHSH form of Bell-inequalities \cite{CHSH}. Besides that, they 
can be seen in terms of closing the 
locality loophole.
Beyond examining fundamental questions, our experiments establish 
also the feasibility of quantum 
cryptography with photon pairs as proposed by Ekert \cite{ekert91} 
over a significant distance. Two short 
paragraphs will briefly discuss our experiments from these two points 
of view (section V). Finally, a brief 
conclusion is given in section VI.

\section{Theoretical background}

\subsection{Tests of Bell inequalities}

A Bell-type experiment consists of the following parts: A source 
emitting pairs of correlated particles,
propagating in different directions.
Each particle enters an apparatus analyzing the correlated feature 
and ascribing a binary value $(\pm 1)$ to 
the outcome. The operation of each device is controlled by a knob 
which sets the parameters $\delta_1 
(\delta_2)$, e.g.  phase shifts of interferometers (when dealing 
with energy-time correlations) 
or orientation of polarizers (when using polarization correlations). 
The classical
information about the detection of a particle, namely when and where 
it is detected, is then send to a coincidence 
electronic which
measures the number of time-correlated events $R_{i,j}(\delta_1,\delta_2), 
(i,j = \pm1)$. 
$R_{+-}$ denotes e.g. the coincidence count rate between the + labeled 
detector at apparatus
1 and the - labeled one at apparatus 2. This enables to calculate the 
so called correlation coefficient   
\cite {aspect82}

\begin{equation}
E(\delta_1,\delta_2) := \frac{R_{++}(\delta_1,\delta_2) - 
R_{+-}(\delta_1,\delta_2) - 
R_{-+}(\delta_1,\delta_2) + R_{--}(\delta_1,\delta_2)}
{R_{++}(\delta_1,\delta_2) + R_{+-}(\delta_1,\delta_2) + 
R_{-+}(\delta_1,\delta_2) + R_{--}(\delta_1,\delta_2)}
\label{correlationcoefficient}
\end{equation}
\noindent
To evaluate this coefficient from measured data, we have to 
assume that the actually detected particle 
pairs form a representative sample of all created pairs. 
The famous Bell-inequalities point out an upper limit for a combination 
of four such correlation 
coefficients with different analyzer settings $\delta_1,\delta_2$ under 
assumption of local hidden variable 
theories (LHVT). One of the most often used form, known as the 
Clauser-Horne-Shimony-Holt (CHSH)
Bell-inequality \cite{CHSH} is 

\begin{equation}
S = \left|E (d_1, d_2) + E (d_1, d_2') + E (d_1', d_2) - 
E (d_1', d_2')\right| \leq 2,
\label{Bell}
\end{equation}
\noindent
where $d_i$,$d_i'$ (i = 1,2) denote values of phases $\delta_i$.

We now describe the quantum mechanical predictions for a test of 
Bell-inequalities using energy-time 
entangled photons as proposed by Franson in 1989 \cite{franson89}. 
Each one of the two 
entangled photons 
is directed into an unbalanced interferometer.
Since the path-length difference of the interferometers, exactly the 
same for both of them, 
is much greater than the coherence length of the single photons, no single 
photon interference 
can be observed. With reference to experiments using polarization entangled 
photons, we refer to this as 
rotational invariance \cite{CH}. 
However, since two of the four processes leading to a coincidence detection 
(each photon can choose either the short or the long arm of the 
interferometers) 
are indistinguishable, 
fringes can be observed in the rate of coincidence detections between two 
detectors belonging 
to different interferometers. 
Due to the two 
noninterfering possibilities (the photons choose different arms) the 
visibility of the 
interference fringes is limited to 
50\%. However, the latter events can be excluded from registration 
provided the 
detection-time jitter is smaller than the time difference between 
passing through the long or the 
short arm. The coincidences can then be resolved into two satellite 
peaks showing no interference effects 
and a central interference peak (see Fig. 2). Confining counting only 
to events in the middle peak 
\cite{brendel91},
an entangled state is created 
where either both photons pass through the short arms or both through the 
long arms:

\begin{equation}
|\psi\rangle = \frac{1}{\sqrt2}\bigg(|s\rangle_1|s\rangle_2+
|l\rangle_1|l\rangle_2\bigg).
\label{entanglement}
\end{equation}
\noindent
The two processes 
remain coherent with each other if the coherence length of the pump laser is 
longer than the difference 
between short and long arms of the interferometers. The maximum visibility 
can be
increased in principle up to 100\%.  
The quantum mechanical description leads via the coincidence function 

\begin{equation}
R_{i,j}^{QM}(\delta_1,\delta_2) = m(1 +ij V \cos(\delta_1+\delta_2))
\label{coincidencerate}
\end{equation}
\noindent 
($m$ being the mean value and i,j = $\pm$1) to the correlation function

\begin{equation}
E^{QM}(\delta_1,\delta_2) = V \cos(\delta_1+\delta_2).
\label{qm}
\end{equation}
\noindent 
V denotes the visibility, describing experimental deviation 
from the maximum value V=1.
Using Eq. \ref{qm}, the settings 
\begin{equation}
d_1=-\pi/4, d_1' = \pi/4, d_2=0, d_2'=\pi/2,
\label{settings}
\end{equation}
\noindent
and assuming V = 1, Eq. (\ref{Bell}) yields 

\begin{equation}
S= 2 \sqrt{2}.   
\label{violation}
\end{equation}
\noindent
This value is higher than the one predicted by LHVT. The violation thus
shows that the description of nature as provided by quantum mechanics is 
unreconcilable with the 
assumptions leading to Bell-inequalities.

Another type of Bell-inequality was given by Clauser and Horne \cite{CH} 
for an 
experiment with polarizers. A similar argument can 
be applied to experiments using interferometers:
if it is found experimentally that the single count rates are constant, 
and that 
$E(\delta_1,\delta_2)=E(\Delta)$ holds where $\Delta=|\delta_1+\delta_2|$ is 
the sum of the phases in both interferometers , then Eq. \ref{Bell} 
reduces to

\begin{equation}
S = |3E (\Delta) - E (3\Delta)|  \leq 2
\label{Bellsum}
\end{equation}
\noindent
Beyond that, if it is found that the correlation coefficient E is 
described by a sinusoidal function of the 
form (\ref{qm}), then Eq. \ref{Bell} reduces to
\begin{equation}
S = \frac{4}{\sqrt{2}}V\leq 2.
\label{bellvisibility}
\end{equation}
Hence, observing a visibility V greater than 
\begin{equation}
V \geq \frac{1}{\sqrt{2}}\approx 0.71
\label{violvisibility}
\end{equation}
\noindent
will directly show that the correlations under test can not be 
explained by LHVT.

\subsection{Quantum key distribution}

As pointed out by Ekert in 1991\cite{ekert91},  
two-particle entanglement can be exploited for quantum key distribution. 
The same 
correlations used to show that nature can not be explained by LHVT can 
be used
to establish a sequence of correlated bits between two users, usually 
called Alice and Bob. 
Moreover, a calculation of the Bell parameter S permits them to check 
whether a third, unlegitimate party (usually called Eve) tried to extract 
information
from the quantum channel.
In 1992, two further protocols, similar to the so-called BB84 or 4-states 
protocol
\cite{BB84} and B92 or 2-states protocol \cite{B92}, resp., but based also 
on quantum correlation 
of photon pairs have been published \cite{bennett92,ekert92}. 
The setups are similar to the one already described to test 
Bell inequalities. We will describe only the latter proposal \cite{ekert92}.
 A source emits pairs of entangled particles flying back to back 
towards Alice and Bob. Both have an interferometer to their disposal. 
(We restrain 
ourself to energy time entangled photons but every kind of two particle 
entanglement will do as 
well.) For each incoming photon, Alice randomly choses phases of either 
$d_A$=0 or 
$d_A'$=$\pi$/2, Bob randomly applies either $d_B$=0 or $d_B'$=-$\pi$/2. 
After a series of EPR particles has been measured,
they announce publicly the settings of their analyzers but not which detector 
registered the photon. They then discard all measurements in which 
$\delta_A+\delta_B$ $\neq$ 0 as well as the instances where either or both 
of them failed 
to register the photon. For the remaining instances ($\delta_A+\delta_B$ = 0) 
the results of their 
measurements should be perfectly correlated. To assess the security of 
their communication,

Alice and Bob publicly compare a random part of their key. If they find 
that the tested bits 
are perfectly correlated, they can infer that the remaining bits are also 
perfectly correlated 
and that no eavesdropper has tried to listen in. The remaining bits can 
now be used to form the
cryptographic key.

In practice, 
also if no eavesdropper disturbed the key exchange, 
there will always be some corrupted bits due to 
imperfections of the experimental setup. Using standard error correction 
schemes, they can be localized 
and removed. However, since Alice and Bob can never be sure whether the 
presence of uncorrelated bits are due 
to the poor performance of their setup, they always have to assume that 
all errors are caused by an 
unlegitimized third person. The 
information about the key that Eve might have gained can be reduced 
arbitrarily 
close to zero using a procedure called privacy amplification 
\cite{privacyamplification}. 
However, this procedure only works if the 
common information between Alice and Bob is higher than the one between 
Eve and any of the two others. This 
is the case whenever the bits, Alice and Bob share, remain sufficiently 
well correlated in order to 
violate Bell inequalities \cite{EveBell}.

It is interesting to note that besides ensuring the security of entanglement 
based quantum cryptography, the Bell inequality 
is even connected to the one qubit application of quantum cryptography: an 
eavesdropper (Eve) on a
quantum channel can get more information than the receiver (Bob)
if and only if the noise she necessarily introduces in the channel by 
eavesdropping is so
large that Bell inequality can no longer be violated \cite{EveBell}. 

\section{Experimental setup}

\subsection{General setup}

The schematic setup of the experiment is given in Fig.~1. 
A source creating pairs of energy-time entangled photons is placed at a 
telecommunication 
station near Geneva downtown. One of the correlated 
photons travels through 8.1 km of installed standard telecom fiber to an 
analyzer that is 
located in a second station in 
Bellevue, a little village 4.5 km north of Geneva. Using 
another installed fiber of 9.3 km, we send the other 
photon to a second analyzer, situated in a third station in Bernex, 
another little village
7.3 km southwest of Geneva and 10.9 km from Bellevue. Absorption in 
the connecting fibers are 5.6 dB and 
4.9 dB, respectively, leading to overall 
losses in coincidences of about a factor of 10. The analyzers consist 
of all fiber-optical interferometers 
with equal path length differences. Behind the interferometers, the 
photons are detected by photon counters 
and the (now classical) signals are transmitted back to the source where 
the coincidence electronics is 
located. Finally, the results 
of the measurements made at different analyzers are compared in order to 
reveal the nonlocal correlations.

\subsection{The two-photon source}

The main elements forming a two-photon source are a pump laser and a 
nonlinear crystal (see also Fig ~1). 
To generate pairs of energy-time entangled photons suitable for long 
fiber transmissions, 
the pump laser has to have the following properties: Its wavelength 
must be adjustable  
in order to create photon pairs at a wavelength 
at which losses and pulse broadening caused by chromatic dispersion 
(see section dispersion in fibers) are small. 
In order 
to work in the second telecommunications window at 1310 nm, the 
wavelength of the pump laser (half the 
wavelength of the created photon pairs) should thus be tunable 
around 655 nm. 
Besides that, its 
coherence length must be large compared to the path-length difference 
short-long of the interferometers in 
order to maintain the coherence of the processes   
$|s\rangle_1|s\rangle_2$ and $|l\rangle_1|l\rangle_2$. Comercially available 
laser diodes more or less meet these requirements. Their wavelength can be 
slightly tuned 
by changing their 
temperature. For example, a drop of 5$^\circ$C will go along with a decrease 
of wavelength of  
$\approx$ 1 nm. The 
coherence length of such diodes varies with temperature and laser current 
and can attain values of up to 
50 cm. In our 1997 experiment, we used a laser diode from RLT 
(6515G; 8 mW at 655.7 nm). 
The coherence length was long enough to demonstrate the existence of the 
entangled state 
(Eq. \ref{entanglement}). However, in order to use energy-time entangled 
pairs for applications like quantum key 
distribution, a better performance is necessary.  
The two-photon source used in our 1998 experiment was based on a  
laser diode with external cavity (Sacher Lasertechnik; 10~mW at 654.8~nm) 
having a coherence length 
of around 100 m.  

The light from the pump laser passes through a dispersion prism P to 
separate out the 
residual infrared fluorescence light and is focused into a 
$\mbox{KNbO}_3$ crystal (Casix) (see Fig.1). The crystal is 
oriented to ensure degenerate collinear type-I phase matching for 
signal and 
idler photons at  1310~nm (hence the downconverted photons are both 
polarized orthogonally with respect to 
the pump photon).
Due to these phase-matching conditions, the single photons exhibit 
rather large bandwidths of 
about 70~nm full width at half maximum (FWHM). Behind the crystal, the 
pump light is separated out by a 
filter F (RG~1000) while the passing down-converted 
photons are focused (lens L) into one input port of a standard 3-dB 
fiber coupler. 
Therefore half of the pairs are split and exit the source by different 
output fibers. The whole source including stabilization of laser current 
and temperature is of small 
dimensions and hence can easily be used outside the laboratory 
(in  1997 a box of 
about $40\times45\times15$ $cm^3$, in 1998 two boxes, each of 
about $30\times40\times15$ $cm^3$).

\subsection{The interferometers}

The two analyzers consist of all-fiber optical Michelson interferometers 
made of standard 3 dB fiber 
couplers. In the 1997 experiment we used chemically deposited end mirrors, 
in the 1998 one so called Faraday mirrors FM to reflect the light. (For 
the advantages of the latter 
solution see the section 
about dispersion in optical fibers and ref. \cite{TittelEuroPhysLett}.) 
The optical path-length differences 
(20 cm of optical fiber or 1 ns time difference in the 
1997 experiment, 24 cm of optical fiber or 1.2 ns 
time difference in the 1998 one) are within 10 $\mu$m equal in all 
interferometers. To control and 
change them, the temperature of the devices can be maintained constant 
or can slowly be varied.

To build fiber optical interferometers with almost identical path-length 
differences, we proceeded 
along the following lines (we first give the description for the 1998 
experiment): 
In a first step we set up two interferometers with roughly 8 mm 
difference of 
path-length difference (from now on called discrepancy). In a second step, 
we measured the exact value of 
this discrepancy.
To do so, we connected the two interferometers together with a third 
bulk-optical interferometer in series 
and illuminated them with a LED. 
When scanning the bulk-optical interferometer, one can find interferences 
if the path-length difference in the bulk optical interferometer equals 
the discrepancy within 
interferometers one and two. 
By this means one can measure this discrepancy with a resolution of a 
few $\mu$m. Changing the  
path-length difference of one of the two interferometer by cutting off 
the additional length of fiber, 
we were able to build interferometers with equal (within 10 $\mu$m) 
path-length differences. As we use 
Faraday mirrors to reflect the light, hence can not cut off the end of 
a fiber, we had to chop a 
peace of fiber in the middle of the arm.
The tool used to cut the fiber precisely was made out of a fiber 
cleaver (Fujitsu) and a
micro-translation stage. 
In the 1997 experiment, we directly cut the interferometers to have 
the most similar path-
length differences possible. 
After having measured the discrepancy, we did the final alignment by 
polishing one interferometer arm. 
Only after this procedure the fiber ends were reflection coated. 

In order to have access to the 
second output ports of the analyzers, normally coinciding with the 
input arm for Michelson 
interferometers, we used 3-port 
optical circulators C (JDS Fitel) in the 1998 experiment. 
This non-reciprocal device, based on the Faraday effect,
enables to direct light from an input port 1 to output 2 and from 
port 2, serving now as input, 
to a third port regardless of the polarization state of the light.

\subsection{Dispersion in optical fibers}

Two problems which have to be faced when using optical fibers are 
chromatic dispersion (CD) and polarization 
mode dispersion (PMD), especially when working with light of large 
bandwidth (in our case around 70 nm FWHM).
For both effects, we have to distinguish between dispersion in the 
fibers connecting the source 
to the analyzers, and dispersion in the fibers forming the short and 
the long arms of the 
interferometers. However, for a 
good performance, both effects have to be seen in connection. We will 
first discuss the effects of 
chromatic dispersion. 

In general, a light pulse travelling in dispersive media becomes broadened. 
In our case of coincidence detection of two photons, the increasing 
detection time jitter 
lead to a loss of temporal correlation between the two photons forming 
a pair. The resulting 
less perfect discrimination of the satellite coincidence peaks (see Fig.~2) 
then causes 
a smaller visibility. 
This problem could be solved using interferometers with larger path-length 
differences. However, different pulse broadening in the intereferometer 
arms, increasing with 
increasing arm-length difference, leads to a smaller visibility as well. 
It has been shown 
\cite{franson92,steinberg92,larchuk95,brendel98} that each of both CD 
effects, inside and outside the interferometer, can be cancelled out 
using photon pairs 
created by parametric downconversion. 
The strict anti-correlation of signal and idler photon enables to 
achieve a dispersion for one photon 
which is equal in magnitude but
opposite in sign to that of the sister photon. The effect of broadening 
of 
the two wave packets then exactly wipes out (assuming a linear dependence 
of CD in function
of the optical frequency, a realistic assumption). 
Two incidentally coincident photons stay coincident 
and no decrease of visibility due to different wave-packet broadening occurs. 
However this cancellation 
requires a choice of the frequencies of signal and idler photon which 
is determined by the dispersion 
properties of the used fibers. As the fibers connecting source and 
interferometers and the fibers inside the 
interferometers are certainly not identical, complete cancellation of 
all dispersion effects at the 
same time is 
impossible. For instance, in the 1998 experiment we used a pump wavelength 
of 654.8 nm to create 
photon pairs at 1309.6 nm, 
leading to a dispersion caused time jitter which is below our limit of 
resolution, hence does not prevent 
from
discriminating the satellite peaks. (The zero of CD for the fiber going 
to Bellevue is around 1312 nm, 
for the fiber going to Bernex around 1305 nm.) With two classical light 
pulses, this cancellation effect would be 
impossible. 
Even in the best case of centering the pulses around the zero 
of chromatic dispersion, we estimate a CD caused broadening of at least 
600 ps for pulses of the 
same bandwidth.
It is difficult to estimate the limitation of the 
visibility due to chromatic dispersion effects in the fibers forming 
the interferometers
as we do not know the dispersion data of those fibers. However, even a 
deviation of a few nm from the zero dispersion wavelength (usually close 
to 1310 nm) 
causes only small effects. Therefore the influence on the visibility 
should be neglectable.

We now discuss the problems caused by PMD, first the effect of 
depolarization in the connecting fibers 
\cite{PMD}. 
Indeed, if the relative delay $\Delta\tau$ between the two modes of 
polarization transmitted by an optical 
fiber is larger than the coherence time of the photons 
(please note that the so called single mode fiber actually guides two 
modes of polarization), then an 
initially polarized photon completely looses its polarization.  
The single photons transmitted in our experiment have a bandwidth of 
approximately 70 nm (FWHM), corresponding to 
a coherence time $\tau_c$ of around 90 fsec $(FW\frac{1}{e}Max)$.
Using the standard value for PMD for modern telecommunication fibers of 
0.5 ps/$\sqrt{km}$, 
and a fiber length of 8 km, we find 
$\Delta\tau = 0.5 \frac{ps}{\sqrt{km}}\sqrt{8 km} \approx 1.4 ~ ps$
which is more than one order of magnitude larger 
than $\tau_c$. Thus the photons arriving at the stations in Bellevue 
and Bernex, respectively, 
are completely depolarized, a result we confirmed in a seperate experiment. 
Hence, an experiment using  
polarization entangled photons with similar bandwidth would be impossible. 
However, as our experiment does not take advantage of polarization 
entanglement, the effect of PMD in     
the connecting fibers is without consequences. In contrast to that, 
PMD in the fibers forming the interferometers either  
has to be avoided or to be compensated as it will lead to a decrease 
of visibility. 
In the 1997 experiment, we aimed to avoid all birefringence by placing 
the fibers which form the arms of the 
interferometer straight and without stress into copper tubes. However, 
a small temperature dependent 
birefringence could still 
be observed, probably caused by mechanical stress induced by the housing 
of the fiber coupler. To overcome 
this inconvenience, we used so called Faraday 
mirrors in the 1998 experiment \cite{Faradaymirror}. This device consists 
of a $45^\circ$  Faraday rotator in front of a 
conventional mirror 
to reflect the light at the end of the interferometer arms. These mirrors 
ensure that a photon, 
injected in any arbitrary 
polarization state into one of the interferometric arm will always come 
back exactly 
orthogonally polarized, regardless any birefringence effects in the fibers. 
Hence, 
no polarization alignment is required.

\subsection{The photon detectors}

To detect the photons, we use germanium avalanche photodiodes 
(APD; NEC NDL5131P1) which we operate in the so 
called Geiger mode. This means that the bias voltage exceeds the breakdown 
voltage, leading an impinging 
photon 
to trigger an electron avalanche which then causes a macroscopic current 
pulse. After detection of this 
pulse, 
the avalanche has to be stopped and the diode to be charged again. A large 
(typically 50 k$\Omega$) 
resistor is connected in series with the APD. This causes a decrease of 
voltage across the APD below 
breakdown after the beginning of an avalange, and thus leads to so called 
passive quenching of the 
avalanche. The recover time of the diode is given by the value of the quench 
resistor and their capacity. 
The emission of electrons trapped during the process of recharging leads to 
an elevated possibility to get 
a count not caused by a photon after an avalanche has taken place. The 
afterpulse fraction describes the 
probability to count such an afterpulse. For a more thorough review of 
photon counting with APDs, which is 
beyond the scope of this paper, we refer the interested reader to 
\cite {GeAPD}.

Unfortunately, germanium APDs show a 
lot of dark counts D, a higher afterpulse fraction and a smaller 
efficiency $\eta$ 
compared to silicon APDs which can be used 
to count photons only of up to about 1000 nm wavelength.
Hence, the advantage to use a wavelength 
where fiber losses are low in order to achieve long transmission 
distances has a drawback: the
possibility to have an accidental coincidence caused by two dark 
counts happening at the same time or by a  
detection of a photon simultaneously with a dark count instead of 
the correlated photon which 
has been absorbed
is high compared to short distance experiments using silicon detectors. 
Thus, the true
coincidences might be hidden behind accidentals.
The maximum achievable visibility for photon pair interference without 
subtraction of accidental 
coincidences is limited by the number of detected 
coincidences in the interference maximum (C) and the number of 
accidental ones (A). 
\begin{equation}
V_{max} = \frac{C - A}{C + A}.
\end{equation}
Hence, to achieve a visibility above 0.71, the ratio of detected 
to accidental coincidences (C/A) has to be larger than 
\begin{equation}
C/A_{critic} = 5.9,
\label{five}
\end{equation}
\noindent
assuming no reduction of visibility due to any other causes.

As usual, we operate the APDs at liquid nitrogen temperature (77K) in 
order to decrease the number of 
dark counts. 
To quench the avalanches we use a relatively high resistor of 180 k$\Omega$. 
The 
long recover time guarantees that most of the trapping centers are already 
empty before the 
diode is charged 
again, hence ensures a low afterpulse fraction. At the same time, the 
quantum-efficiency-to-noise ratio 
$\eta /D$ increases. Besides the high quench resistor, in the 1998 
experiment we implement large
electronic deadtimes of about 4 $\mu$s. By this means we suppress the 
counting of pulses when the diode is 
not completely charged, which would lead to an increasing time-jitter. 
In the 1997 experiment, a quench 
resistor of only 50 k$\Omega$ had been chosen and no extra electronic 
deadtime was applied. However, the 
performance of the subsequent time to pulse hight converter 
(see section V.G.) ensures a similar deadtime, 
at least for the detector providing the start pulse.

To ensure that the overall quantum efficiencies in both detectors 
attached to the same interferometer 
are equal, we adjusted bias voltage and additional losses for both 
detectors in a way, that dark and light 
count rates are as similar as possible. We operated the detectors within 
a regime where  
dark count rates are of roughly 25 kHz, and we find quantum efficiencies 
of about 5$\%$. 
The setup of our two-photon source leads to a separation of 50 $\%$
of the created photon pairs. 
Losses in the connecting fibers are 90 $\%$
and excess losses in each interferometer around 50$\%$. In addition, 
we loose 50 $\%$ of 
coincidences due to a small coincidence window. Finally, and
being a fundamental problem for Franson type experiments, the 
discrimination of the satellite coincidence peaks further reduces 
the coicidences by a factor of two. 
All together, we find a probability to detect an emitted photon 
pair of about 8*$10^{-6}$.
The time jitter for the coincidence detection is around 350 ps 
FWHM and 
ensures a negligible contribution of the satellite coincidence peaks. 
We measure a ratio C/A of around 
20 which permits to violate Bell inequalities without subtraction of 
accidental coincidences.
Using the diodes at lower dark count rates would 
increase the fraction $\eta /D$ even more, however, the growing time 
jitter would require a 
larger coincidence window, hence would lower the ratio C/A. Moreover 
a more important part of the 
satellite peaks would fall into the window as well.

\subsection{Transmission of results of measurements}

The classical information about detection time and detector number 
has to be transmitted to a common place (in our 
case the place where the source is located). To do so, we use
supplementary telecommunication fibers. The two possibilities to 
detect the photon (either 
detector + or --) is encoded in series of two short laser pulses 
separated by a short (detector --) or a long 
(detector +) delay. The pulses are detected by ordinary pin photodiodes and 
the delay between the pulses is transformed into the detector label again. 
Another possibility for 
transmission of the classical information would be to use one fiber for 
each detector. However the latter
solution would bring up the need for additional pulsed lasers and 
pin photodiodes. 

Care must be taken not to 
introduce additional time jitter during the processes of coding, 
transmission and decoding, as a large 
incertitude on the arrival time will lead to a loss of temporal 
coherence and hence to a superposition of 
the satellite and the central coincidence peaks.

\subsection{Measurement of correlations}

In order to reveal the nonlocal correlations, one has to compare 
the results of the measurements at the 
distant analyzers. 
The signals from the pin photodiodes trigger time to pulse height 
converters (TPHC; Tenelec ?). We 
choose the signals coming from Bellevue to start, the signals coming 
from Bernex to stop the TPHCs.
For each 
pairing of detectors belonging to different interferometers, we get 
a series of three peaks in the time 
spectrum (Fig 2). Window discriminators permit to count coincidences 
within intervals 
of a few hundred ps which are centered around the interference peaks. 
We measure the four different coincidence count rates 
$R_{i,j}$
in a single run, yielding directly the correlation coefficient 
E($\delta_1, \delta_2$) 
(Eq. \ref{correlationcoefficient}) (in the 1997 experiment, only 
one coincidence function 
was measured, hence the correlation function had to be deduced 
from symmetry arguments).

Please note that it is important to register each pairing of detectors 
with a different 
TPHC. Indeed, separating coincidence peaks belonging to different couples 
of detectors only by introducing 
different delays between start and stop leads to summation of the 
accidental coincidences of all pairings 
and thus to a decrease of the ratio C/A. As we used only two TPHC, 
we utilized a kind of multiplexing in order 
to register all coincidence count rates. Each TPHC was triggered by 
two different pairings of 
detectors. By assigning each TPHC output to the belonging pairing, 
we could overcome the problem of 
counting the ensemble of accidental coincidences in each channel.

\section{Results}

We monitored the four coincidence count rates as a function of 
time and slowly changed the 
phases $\delta_1, \delta_2$ while measuring. 
As the coherence length of the single photons is five orders of 
magnitude smaller than the arm-length 
difference of the interferometers, no phase dependent variation 
of the single count rates can be observed.
Hence our assumption of rotational invariance is well satisfied. 
However, the coincidence count rates 
as well as the correlation coefficient, calculated from the four 
rates using 
Eq. \ref{correlationcoefficient}, show sinusoidal variation when 
changing the 
phases in the interferometers.

\subsection{Experiments with two interferometers}

In order to test the quantum mechanical predictions that the 
correlation function depends only on the 
sum of the phases in both 
interferometers and not on the actual phases in either one, we 
perform the following experiment. We 
change the path-differences of both interferometers first in 
opposite directions, then in the 
same directions and compare the frequencies observed for the 
correlation function with the frequencies 
measured when scanning only one of the two interferometers 
(Fig.~3, table I, table II).
Calculating the frequencies for a joint scan of both interferometers 
from the frequencies 
observed when scanning only one, we find them to be in almost perfect 
accordance with the 
measured values. From this, we can conclude
that indeed the correlation function does depend on the sum of the 
phases in both interferometers 
$(\delta_1+\delta_2)$ 
as described by Eq. \ref{qm}. Hence we can calculate the parameter S 
from the observed visibilities 
(Eq. \ref{bellvisibility}).

In all cases we systematically find values exceeding the limit given 
by the Bell-inequalities
by at least 8 standard deviations ($\sigma$).  
The raw data for a variation of the 
Bernex-interferometer yield a visibility of (86.2$\pm$1)$\%$, leading to  
$S_{raw}$ = 2.44 and a violation of Eq. \ref{bellvisibility} by
 15.5 $\sigma$. Most
of the difference between this result and the theoretical prediction 
of S=2$\sqrt2 \approx 2.83$ can be
attributed to accidental coincidences.

We measure them by delaying the stop signals by additional 8 ns. 
Therefore the signals representing 
correlated photons arrive apart from the detection window. 
We thus destruct all 
correlation between the signals from the two detectors, leaving only 
accidental coincidences to be measured. 
However, we find the true value  
if and only if there is no elevated coincidence rate caused by detection 
of a photon 8 ns before. 
Since the time spectrum shows an uniformly distributed noise floor 
(the detector deadtimes prevent from counting an afterpulse up to 
4 $\mu s$ after detection of a 
photon), it is natural to assume that we can indeed infer from the 
measured to the true 
rate of accidental coincidences. 
Besides, the measured rate of 26.4$\pm1.3$ per 30 sec 
is in excellent agreement with the one we can calculate from the single 
count rates (39.5 kHz) 
and the size of the coincidence
window ((550$\pm$10) ps). Indeed, by the latter means we find 25.7$\pm0.5$ 
accidental coincidences per 
30 seconds. Subtracting them, we obtain 
$V_{net}=(93.3\pm1.1)\%$, 
corresponding to $S_{net}=2.64$ and a violation of Eq. \ref{bellvisibility} 
by 20.5 $\sigma$.

In two further measurements we changed 
the path-length difference in either one of the two interferometers 
within a quite large range 
while the other interferometer is kept stable. The results are 
listed in table III. 
Fig. 4 shows the variation of the correlation coefficient observed 
for a scan in the 
Bellevue-interferometer. The 
correlation function shows a sinusoidal function with a gaussian 
envelope representing the coherence 
length of the single photons \cite{envelope}. From this envelope, 
we calculate a coherence length 
of around $13 \mu$m corresponding to a bandwidth of around 70 nm FWHM. 
Fitting only the two central periodes of the 
correlation functions with a sinusoidal function, we find visibilities 
of up to $V_{raw}$ = (85.3$\pm$0.9)$\%$ 
 ($V_{net}$ = (95.5$\pm$1)$\%$), 
leading to $S_{raw}$ = 2.41 ($S_{net}$ = 2.70) and a violation of 
Eq. \ref{bellvisibility} 
by 16.2 $\sigma$ (24.8 $\sigma$).

Besides determining the correlation functions for the above mentioned 
measurements, we made fits 
of the underlying coincidence functions $R_{i,j}$ (i,j=$\pm$1) 
as well. The results can be found in tables I and III.
We find the visibilities to be in close agreement with the values 
for the correlation function. However, even if the single count rates 
show almost 
perfect symmetry, there is a difference in the mean values of the 
coincidence counts. We found the cause 
for this 
effect to be the large bandwidth of the single photons in connection 
with different spectral 
quantum efficiencies of the different detectors. 
Therefore the sum of 
the coincidence rates of one detector with both detectors on the other 
side (i.e. $R_{++}$ + $R_{+-}$) 
is not constant. However, if summing over all coincidence rates, we 
always find the 
same value, confirming that the size of 
the detected samples of photon pairs does not change.

\subsection{Experiments with three interferometers}

In order to test the CHSH-Bell inequality (Eq. \ref{Bell}), we have 
to measure the correlation coefficients for the discrete phase 
differences given in Eq. \ref{settings}. 
To do so, we modify our setup in the following way (see inlet in Fig.~1). 
The fiber 
arriving in Bellevue is connected to a standard 3 dB fiber coupler. 
Each output arm of this 
coupler is followed by an interferometer of the same kind as described 
before. Hence each incoming photon 
is analyzed by one of the two different phase settings. 
As we 
did not have enough circulators and detectors, we were able to 
observe only one output of each 
analyzer. For this reason we could only measure two of the four 
coincidence count rates needed to 
calculate the correlation function (Eq. \ref{correlationcoefficient}). 
To infer 
from the measured functions to the correlation function we thus have 
to assume the same 
symmetry between 
the coincidence functions as we 
found in the experiments described before. With this quite natural 
assumption, we can evaluate the 
correlation functions $E(d_1, d_2)$ and $E'(d_1', d_2)$ at the same time, 
hence for exactly the same setting $d_2$. Fig.~5 shows the correlation 
coefficients 
observed when 
changing the phase $\delta_2$ in the Bernex interferometer. We find again  
sinusoidal functions, the parameters for best fits are listed in 
table IV. For the difference 
of phases $\delta_1 - \delta_1'$ between the two interferometers 
in Bellevue we obtain
$\pi/2.25$, which is close to the ideal value of $\pi$/2, needed 
to maximally violate Bell 
inequalities. 
Visibilities are about 77.5$\%$ without and about 95.5 $\%$ with 
subtraction of accidental 
coincidences. We can 
now directly evaluate the value of the Bell parameter S from looking 
at the correlation coefficients 
for two different parameters $\delta_2$. For the indicated points we 
find $S_{raw}$ = 2.38 $\pm 0.16$ and 
$S_{net}$ = 2.92$\pm0.18$ leading to a violation of 2.4 respectively 
5.1 standard deviations. Using the 
parameters obtained for the best fits in order to more precisely 
determine value and incertitude of the four 
points, we find $S_{raw}$ = 2.186 $\pm 0.033$ and 
$S_{net}$ = 2.692$\pm0.038$ leading to a violation of 5.6 respectively 
18.2 standard deviations.

\subsection{Experiments with four interferometers}
      
In order to measure all four different correlation coefficients 
required to test the CHSH-Bell inequality at the same time, we perform 
an experiment with four interferometers, 
one couple on each side of the source. This time, the whole setup is 
located in our laboratory. The 
interferometers are placed 2 meters from the source with connecting 
fibers of 5 meters. Each combination 
of interferometers leads to a different correlation coefficient. 
Again we can measure only 
four different coincidence rates and thus have to 
assume the same symmetry as we already did before. Doing so, we can 
calculate the correlation coefficient by
normalizing the coincidence rates with their mean value. From 
Eq. \ref{coincidencerate} and Eq. \ref{qm} we get 
\begin{equation}
E_k(\delta_1, \delta_2) = \frac{R_k(\delta_1, \delta_2)}{m}-1,
\label{normalization}
\end{equation}
\noindent
k denoting one of the four possible combinations of interferometers. 
We fix the difference between 
phases of one couple of interferometer ($I_3, I_4$) to
be $\pi$/2. Then we scan the two other interferometers with different 
frequencies. This leads to sinusoidal 
variation of all coincidence rates (see Fig~6). Best fits enable to 
very precisely determine mean values, 
frequencies and phases (see table V). Using Eq. \ref{normalization} 
we calculate the correlation 
coefficients $E_{k}$, leading to the Bell-parameters S
\begin{equation}
S = E_1 + E_2 + E_3 - E_4.
\label{S}
\end{equation}
\noindent
The results are shown in Fig~7. Using the values of the four indicated 
points, we find 
$S_{raw}$ = 2.41 ($S_{net}$ = 2.63) and a violation of the 
CHSH-Bell inequality by 5.1 $\sigma$ (7.5 $\sigma$).
The theoretical prediction for a phase difference of $\pi$/2, sinusoidal 
functions and equal visibilities for all correlation functions leads to
\begin{equation}
S_{theo}= -2\sqrt2 V sin\bigg(\frac{t-a_1}{2\omega_1}+
\frac{t-a_2}{2\omega_2}\bigg) sin 
\bigg(\frac{t-a_1}{2\omega_1}-\frac{t-a_2}{2\omega_2}-\frac{\pi}{4}\bigg)
\label{Stheo}
\end{equation}
\noindent
with t being the time in numbers of measurement intervals. 
Using frequencies and phases found by fitting the correlation 
coefficients (Fig 6, table V) and fitting only 
the visibility V, we find $V_{raw} = (86.6 \pm1.1)\%$, 
($V_{net} = (94.2 \pm1.1)\%$) 
leading to $S_{raw}$ = 2.45 ($S_{net}$ = 2.66) and a violation of the 
CHSH-Bell inequality by 14.5 $\sigma$ (22 $\sigma$) (see also table V).

\section{Discussion}

All our experimental results are in good agreement with quantum mechanics. 
The measured 
two photon fringe visibility around 86\% can be almost entirely explained 
by the detectors noise.
And since we found a similar netto visibility of (94.3$\pm$0.5)$\%$ 
in an experiment carried out in our lab, 
one has to conclude that the distance does not affect the nonlocal
aspect of quantum mechanics, at least not for distances up to 10 km.
As already mentioned in the introduction, no experiment up to date could 
close the detection 
loophole. In particular, long-distance experiments will probably never be 
suitable for 
a decisive test as transmission losses will always be to high to allow 
detection of 
more than 66.7$\%$ of the created 
photons \cite{eberhard93}. Below we discuss separately the relevance of 
our results for the
debate on the locality loophole and for applications in quantum cryptography.

\subsection{The locality loophole}

The locality loophole is based on the assumption that, somehow, the settings 
$\delta_1$ and $\delta_2$ of the
analyzers influence the photon pairs emitted by the source. Hence, each 
settings would
analyze differently prepared photons. In order to close this loophole, the
settings should be chosen only after the photons left the source.
Hence, long distance experiments are favourable. 
Ideally, a physicist (or any being enjoying freedom) would make the 
setting's choice.
But in practice, random number generators are used.
In the experiment by Aspect et al. \cite{aspectswitch}, a "periodic" 
random number generator determines 
into which of two analyzers with fixed parameter settings the particles 
are send (for a discussion on its 
randomness see \cite{zeilingercritic}). 
Obviously, our setups with three and with four 
interferometers are quite similar to the one chosen by Aspect,
provided one assumes 
that the photon takes a random choice at the fiber coupler and that this 
choice is not predetermined 
by a hidden variable which "knows" the settings of the analyzers behind 
the coupler. 
Let us briefly elaborate on this
point. A possible objection would be to refute the existence of randomness. 
But if randomness
exists, a "quantum random number generator" would qualify as the best possible 
choice. 
Admittedly, one could argue that it would be preferable that this quantum 
choice is made by
a system independent of the particle under test. This outside random number 
would trigger 
a fast electro-optic switch. In practice such switches (Lithium Niobate 
modulators) have losses
higher than 50\%, hence it would be equally practical to use a passive 
splitter, 
as in our experiment and to turn off the detectors of one of the analyzer! 
Turning off
detectors can certainly not improve the experiment. But
from a logical point of view, the above discussion shows that the locality 
loophole is not
independent of the detection loophole, since for low detection efficiency 
passive spitters
are equivalent to active ones.

\subsection{Quantum key distribution}

Let us turn now to the first promising application of entangled particles 
in the field of quantum 
communication, quantum key distribution (QKD). 
The quantum bit error rate (QBER, the number of wrong bits 
divided by the number of transmitted bits) of this scheme
is related to the visibility $V$ of the coincidence function before removal 
of the accidental coincidences: 
QBER=$\frac{1-V}{2}$. 
Note, that subtracting the 
accidentals is impossible for quantum
cryptography, as there is no way to determine which coincidence counts are 
accidental and
which are due to a photon pair. 
To guarantee the security of the transmission, the visibility of the 
coincidence functions has to be above $\frac{1}{\sqrt2}$, hence 
the quantum bit error rate 
below $\approx$ 15 $\%$. 
Since we achieve raw visibilities of up to 85.2 $\%$ from which we can 
infer to a QBER of 7.4$\%$,
we demonstrate that quantum key distribution with photon pairs is possible, 
even over distances of 
more than 10 kilometers.

Our source and analyzers are easy to transport 
and our setup does not depend on specially manufactured fibers but 
can be installed in every modern singlemode fiber network working at 
1310 nm. Beyond that, it does not 
require active polarization control. Therefore it is very promising for 
practical implementation of QKD, 
not far from existing QKD schemes working with weak pulses \cite{weakpulses}.
However, a fast switching in order to really exchange a key still has to be 
implemented.
This switching can be done either by a phase modulator or, as we did in our 
last experiments by using a 
fiber coupler connected to two interferometers with appropriate phase 
differences. The advantage of such a 
setup is that no fast random 
generator and switching electronic is necessary. However, as the visibility 
and hence the QBER decreases due 
to increasing losses, this setup is in our case limited to around 10 km, 
a distance which is determined by 
the number of created photon pairs, overall losses and detector performance.
A better way to do entanglement-based quantum cryptography 
would be to use a source employing nondegenerate phasematching in order to 
create correlated photons 
of different 
wavelengths, one at 1310 nm, the other one around 900 nm. 
This would allow to use more efficient and 
less noisy silicon photon counting modules to detect the photons of the 
lower wavelength.
To avoid the high transmission losses of photons of this 
wavelength in optical fibers, the interferometer(-s) measuring these 
photons could be placed next 
to the source.
First investigations show that quantum cryptography over tens of kilometers 
should be possible.

\section{Conclusion}
In conclusion, we have reported on experiments demonstrating strong 
two-photon correlations over more than 
10 kilometers.
Provided that our results are not affected by the remaining loopholes, 
we thus can confirm that nature can not 
be described by LHVT. Beyond that, our experiments support the prediction 
of quantum mechanics 
that distance has no effect
on these quantum correlations. The experimental difficulties and possible
solutions have been discussed in length.

The feasibility of long-distance experiments now opens the door for 
several interesting 
possibilities, both in the field of fundamental tests of quantum 
physics, as in the field
of emerging applications of quantum information processing.
Among the latter, let us mention, in addition to quantum cryptography 
which is discussed in
this article, the fascinating possibility of entanglement swapping 
\cite{entanglementswapping}, dense 
coding \cite{densecoding}
and of quantum teleportation 
\cite{teleportation} at large distances.
Among the even more fundamental issues, one interesting possibility is 
to test relativistic nonlocality \cite{RNL}: 
Set one analyzer in motion such that each on analyzer in his own 
inertial frame detects his
photon first. The projection postulate is then difficult to apply, 
if it applies at all
\cite{AharonovAlbert}.

\section*{Acknowledgments}
This work was supported by the Swiss FNRS and Priority Program for Optics,
by the European TMR network "The 
physics of Quantum Information", contr. no. ERBFM-RXCT960087, and by the
Fondation Odier. We like to thank G. Ribordy and T. Herzog for help during 
the experiments as well as 
J. D. Gautier and O. Guinard for technical support.
The access to the telecommunication network was provided by Swisscom and the
circulators by JDS.



\newpage

\begin{figure}[b]
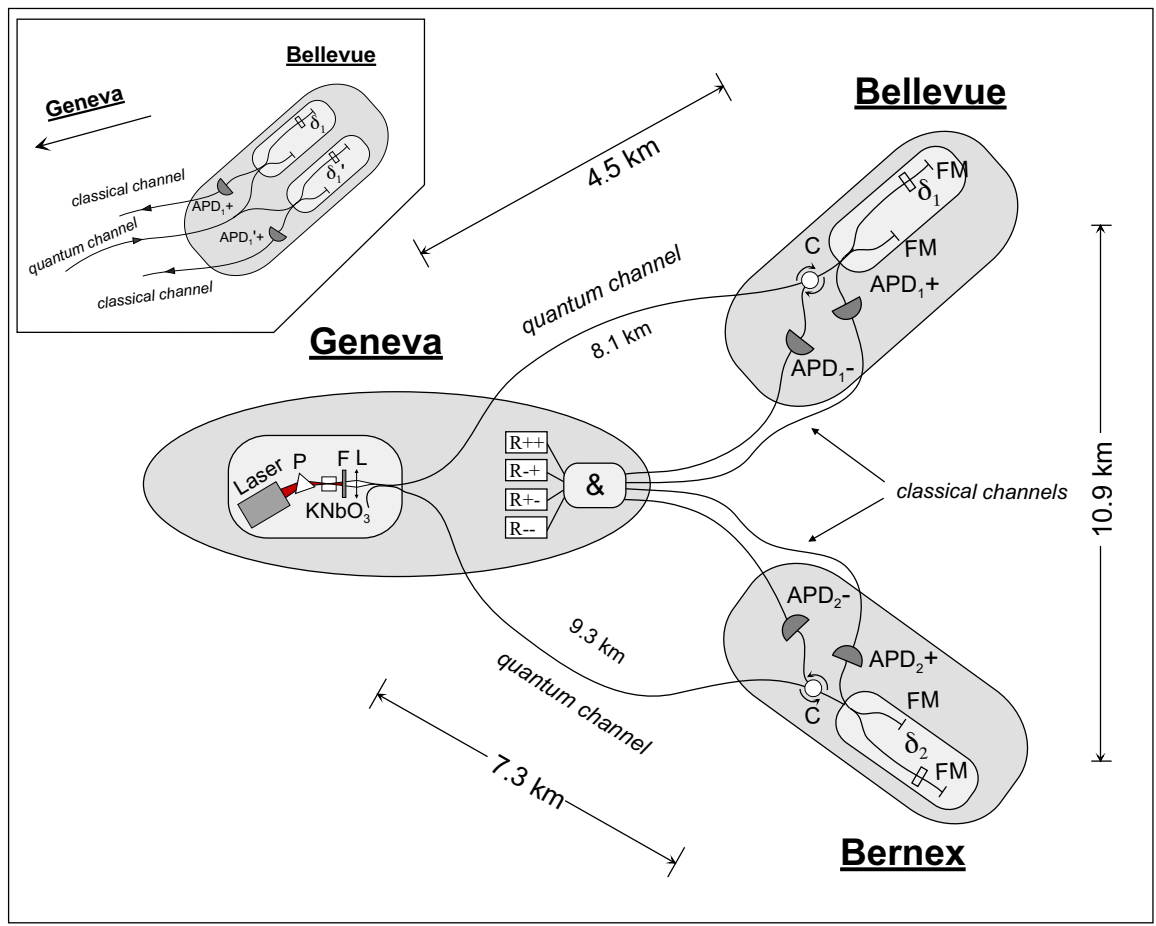

\infig{setup.eps}{0.85\columnwidth}                             
\caption{Setup for experiments with two and three interferometers 
(inlet). See text for detailed 
description.} 
\label{fig1} 
\end{figure}

\begin{figure}[b]
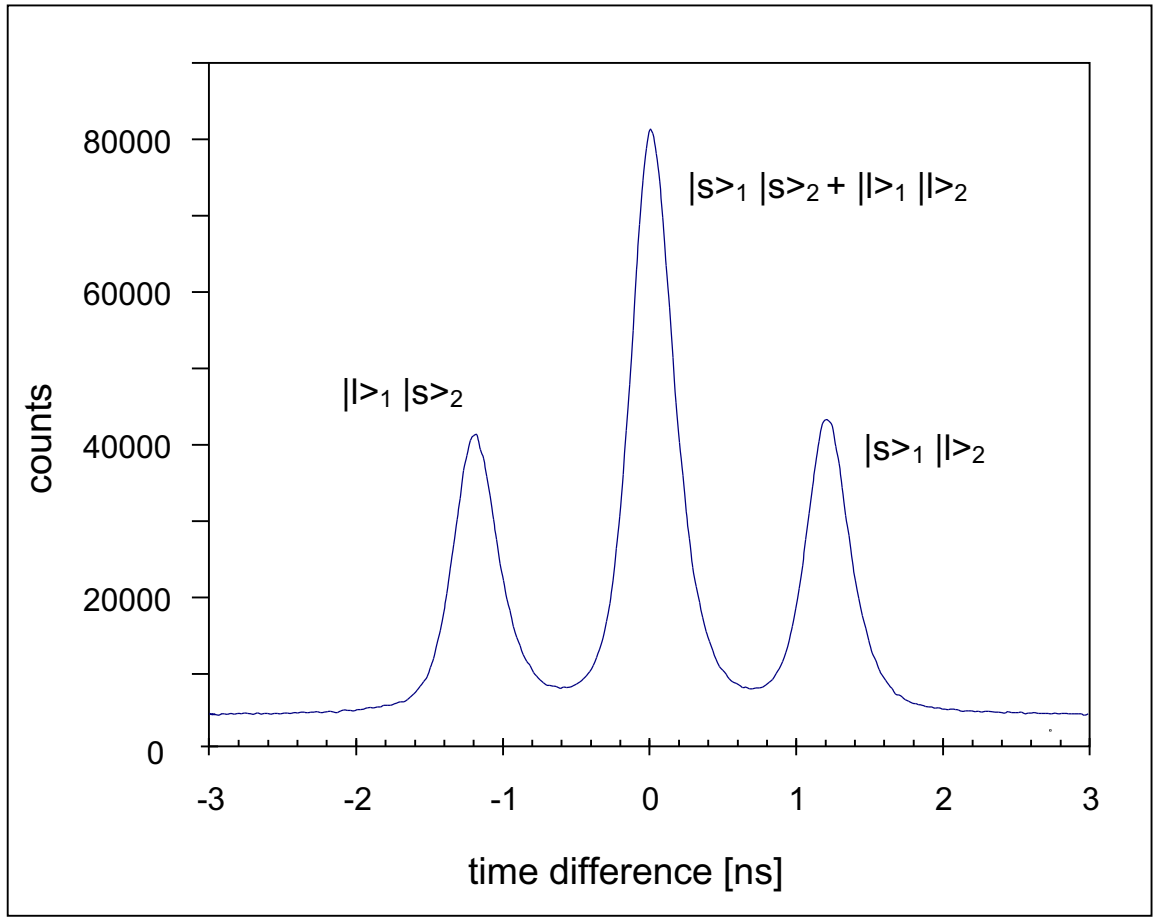

\infig{3peak.eps}{0.85\columnwidth}                             
\caption{Measured distribution of differences in arrival time of 
correlated photons. The sattelite peaks 
belong to the transmission processes long-short (left hand side), 
and short-long (right hand side), 
respectively, in the interferometers. The two 
undistinguishable possibilities short-short and long-long lead to 
a detection within the central peak. Here 
shown is a noninterfering case. The width of the coincidence peaks 
is around 350 ps (FWHM).} 
\label{fig2} 
\end{figure}

\begin{figure}[b]
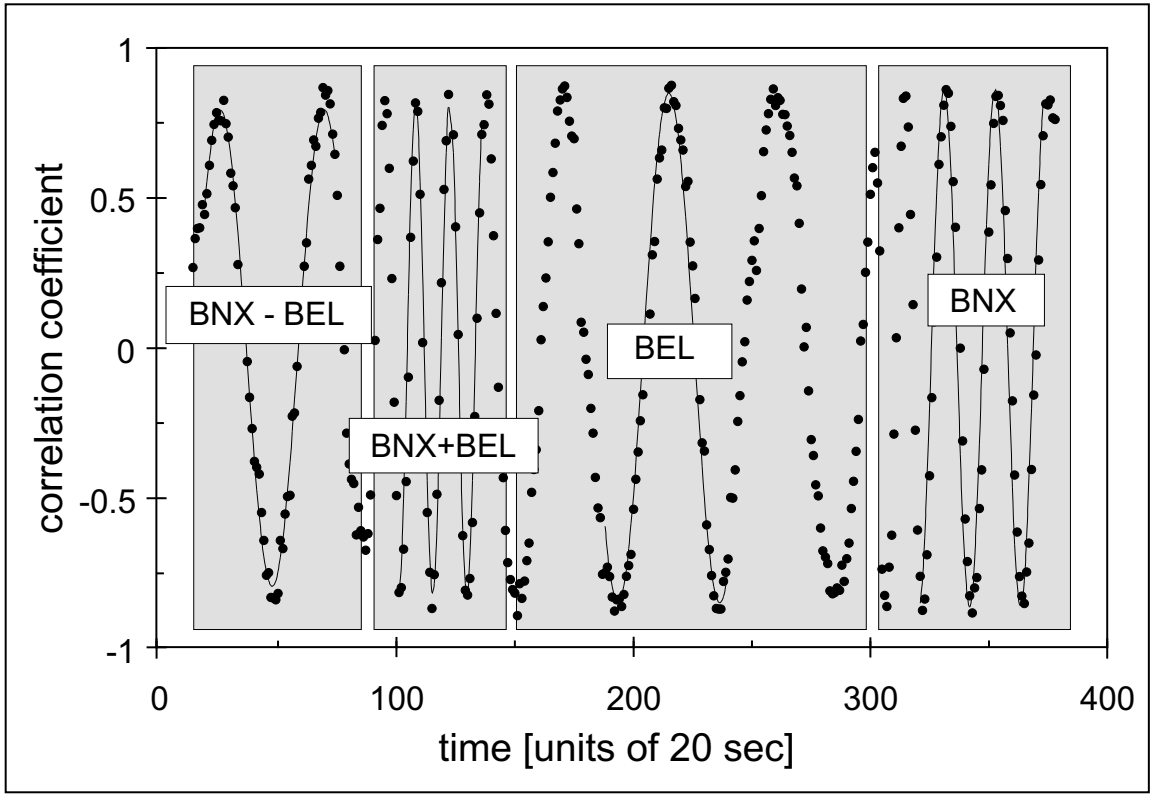

\infig{difsum.eps}{0.85\columnwidth}                             
\caption{Correlation coefficients observed while simultaneously 
changing the phases in both interferometers.} 
\label{fig3} 
\end{figure}

\begin{figure}[b]
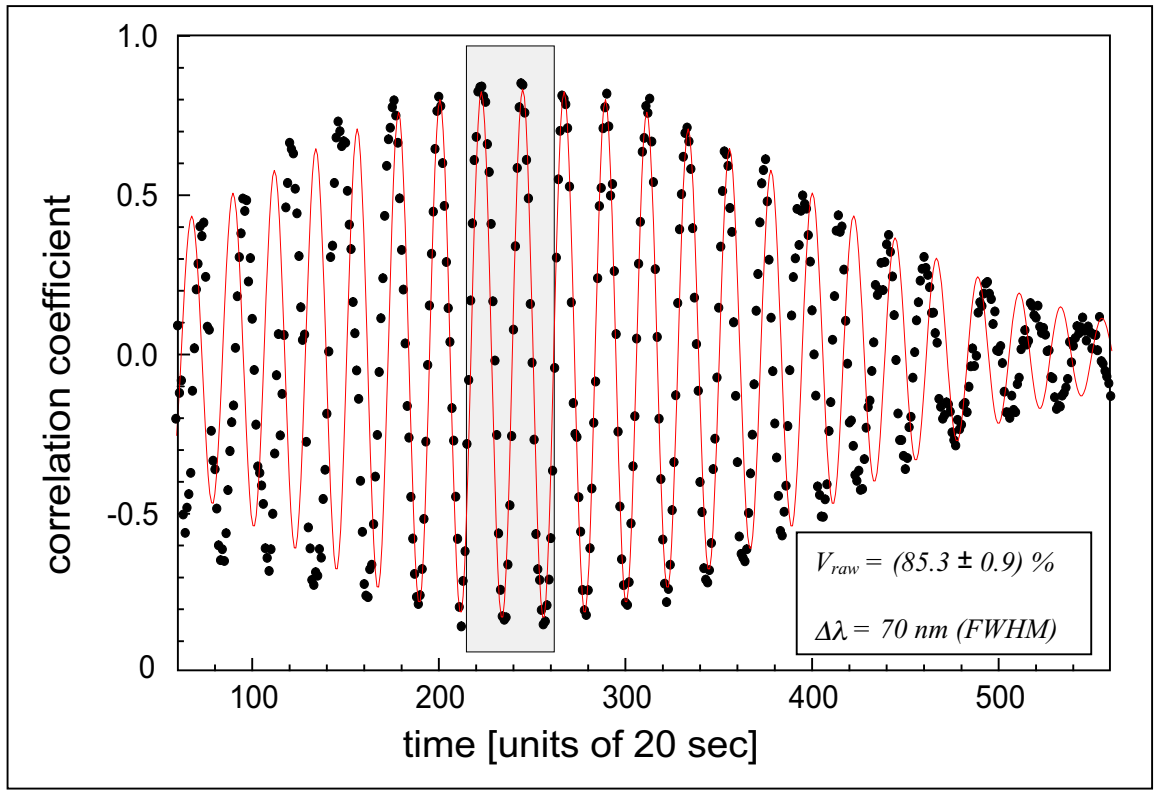

\infig{bellevue.eps}{0.85\columnwidth}                             
\caption{Correlation coefficients measured for large phase-change 
in the Bellevue interferometer. The 
shaded region indicates the region for the fit over two periods.} 
\label{fig4} 
\end{figure}

\begin{figure}[b]
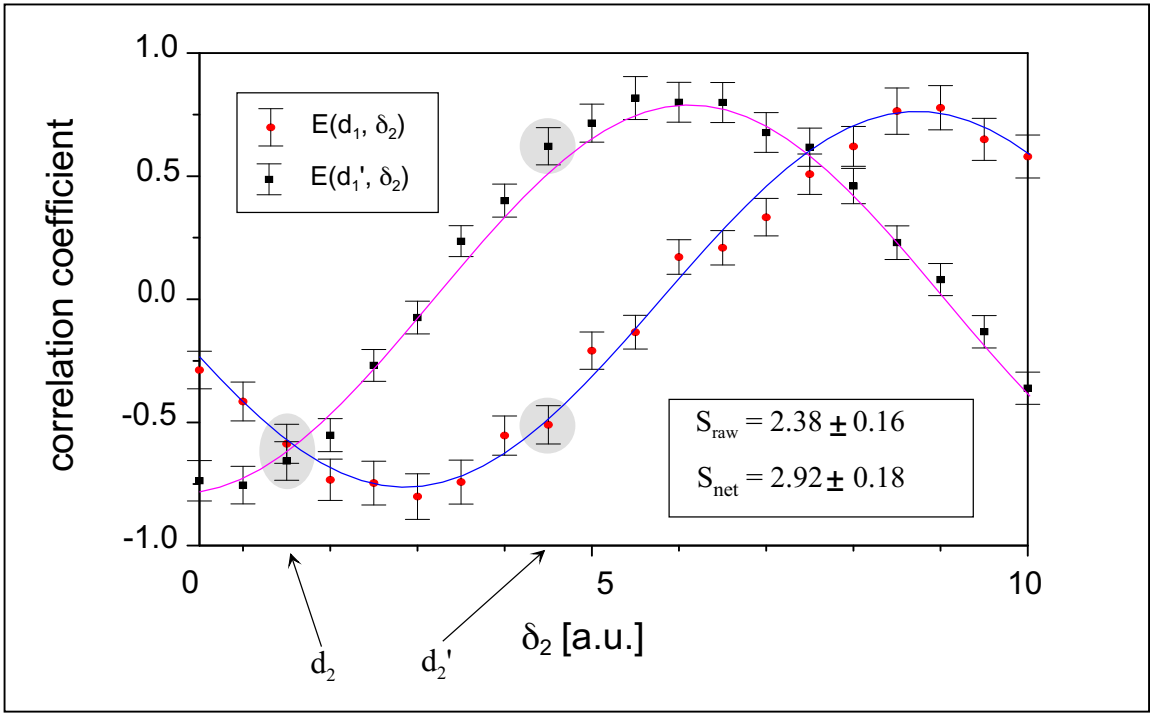

\infig{3int.eps}{0.85\columnwidth}                             
\caption{Correlation coefficients observed in the experiment with 
three interferometers.} 
\label{fig5} 
\end{figure}

\begin{figure}[b]
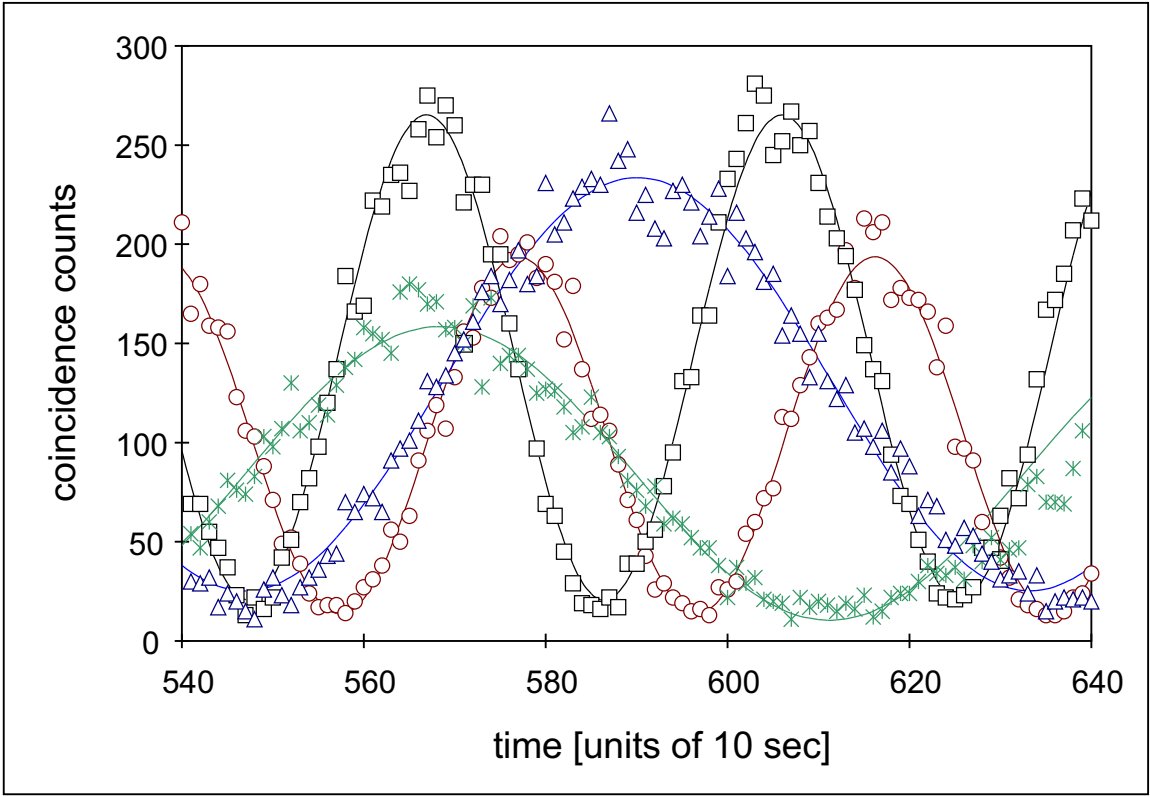

\infig{c_4int.eps}{0.85\columnwidth}                             
\caption{Coincidence count rates observed in the experiment with 
four interferometers.} 
\label{fig6} 
\end{figure}

\begin{figure}[b]
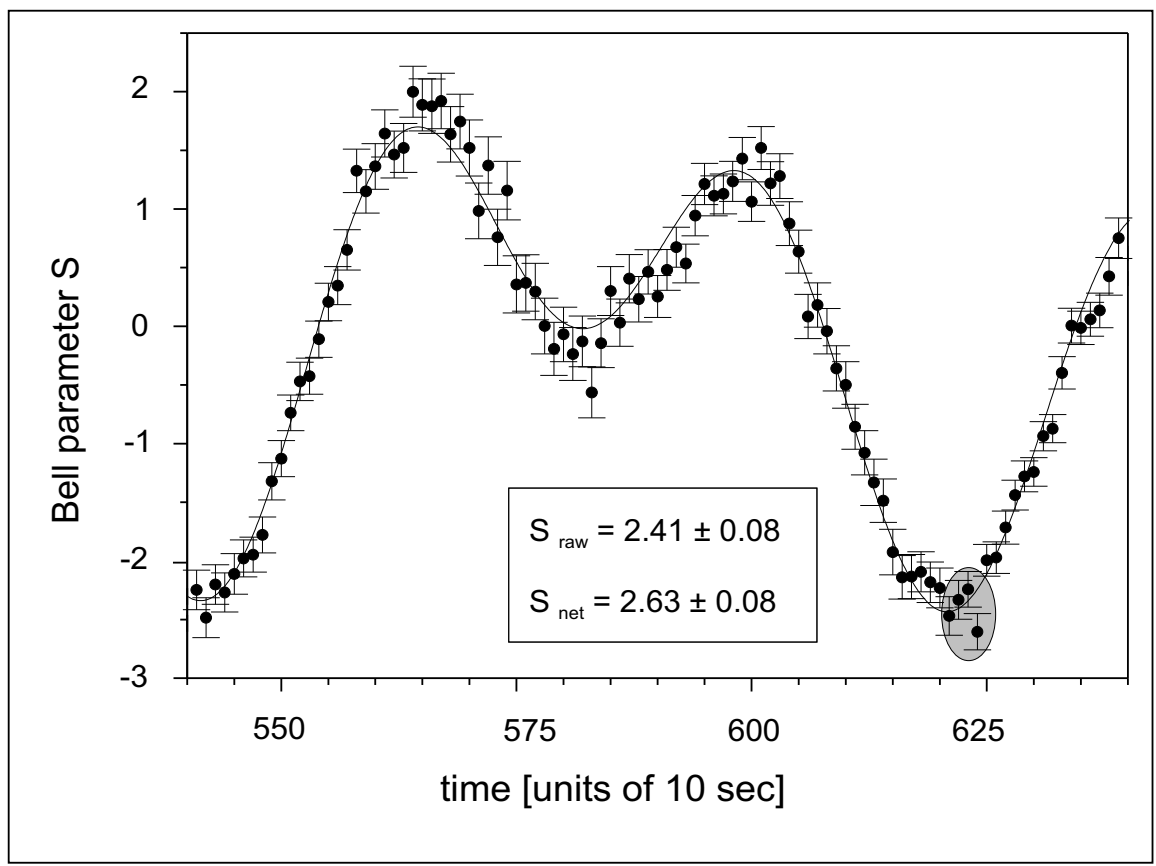

\infig{s_4int.eps}{0.85\columnwidth}                             
\label{fig7}
\caption{Bell parameter S in the experiment with four interferometers.} 
\end{figure} 

\newpage

\begin{table}
\begin{center}
\leavevmode
\epsfxsize=0.85\columnwidth
\newline
\epsffile{table1.eps}
\caption{Results for the experiment with simultaneous phase-change in 
both interferometers (see also table II and Fig. 3).} 
\end{center}
\end{table}

\begin{table}
\begin{center}
\leavevmode
\epsfxsize=0.4\columnwidth
\epsffile{table2.eps}
\newline
\caption{Single count rates, measurement interval and window width 
for the experiment with simultaneous phase-change in both interferometers 
(see also table I and Fig. 3).} 
\end{center}
\end{table}

\begin{table}
\begin{center}
\leavevmode
\epsfxsize=0.85\columnwidth
\epsffile{table3.eps}
\newline
\caption{Results for the experiments with large phase changes in Bellevue 
and Bernex, respectively 
(see also Fig. 4.). The visibility "$l_c$" 
denotes the visibility of the correlation function for a fit over a whole 
coherence length (including the 
gaussian envelope), "2x$\lambda$" the one for a fit over only two
periods. The Bell parameter S is given only for the last mentioned fit.} 
\end{center}
\end{table}

\begin{table}
\begin{center}
\leavevmode
\epsfxsize=0.85\columnwidth
\epsffile{table4.eps}
\newline
\caption{Results for the experiment with three interferometers 
(see also Fig. 5).} 
\end{center}
\end{table}

\begin{table}
\begin{center}
\leavevmode
\epsfxsize=0.85\columnwidth
\epsffile{table5.eps}
\newline
\caption{Results for the experiment with four interferometers 
(see also Fig. 6 and 7).} 
\end{center}
\end{table}


\begin{references}
\bibitem{notthesame}A first attempt to investigate nonlocality over large 
                distances has been reported 
                in \cite{Tapster94}. However, it contrasts with ours as both 
                analyzers where sitting next to the source, one connected 
                with a 4.3 km fiber on a spool, while the other analyzer was 
                directly connected to the source. Therefore the physical 
                distance between the two analyzers was of only a few meters.
\bibitem{envelope}Uequal path-length differences in both interferometers 
        $\delta_1-\delta_2\neq0$ increase 
        the possibility to differentiate between the long-long and 
        short-short paths (by looking at the 
        differences in photon arrival time), hence decrease the visibility.
\end{references}
\end{document}